\documentstyle[epsfig]{aipproc}
\def\lsim{\mathrel{\rlap{\lower4pt\hbox{\hskip1pt$\sim$}}
    \raise1pt\hbox{$<$}}}         
\def\gsim{\mathrel{\rlap{\lower4pt\hbox{\hskip1pt$\sim$}}
    \raise1pt\hbox{$>$}}}         
\newcommand{\beq}{\begin{equation}}
\newcommand{\eeq}{\end{equation}}
\newcommand{\beqa}{\begin{eqnarray}}
\newcommand{\eeqa}{\end{eqnarray}}

\begin{document}

\title{Indirect Detection of Neutralino Dark Matter}

\author{L. Bergstr\"om}

\address{Department of Physics, Stockholm University, Box 6730, 
SE-113 85 Stockholm, Sweden
\\E-mail: lbe@physto.se}


\maketitle

\begin{abstract}
Dark matter detection experiments are getting ever closer to the 
sensitivity needed to detect the primary particle physics candidates 
for nonbaryonic dark matter. 
Indirect detection methods include 
searching for antimatter and gamma rays, in particular gamma ray lines,
in cosmic rays and high-energy neutrinos from the centre of the Earth 
or Sun caused by accretion and annihilation of dark matter particles.
A review is given of recent progress, both on the theoretical and 
experimental sides.\end{abstract}

\section{Introduction}

The mystery of the dark matter is one of the outstanding
questions in standard cosmology \cite{BGbook}. 
One of the  favoured particle dark matter candidates is the lightest
supersymmetric particle $\chi$, assumed to be a neutralino, i.e. a mixture 
of the supersymmetric partners of the photon, the $Z$ and the two neutral
$CP$-even Higgs bosons present in the minimal extension of the 
supersymmetric standard model. The 
attractiveness
of this candidate stems from the fact that its generic couplings
 and mass range
naturally give a relic density in the required range to explain halo 
dark matter. Besides, its 
motivation from particle physics has recently become stronger due to 
the apparent need for 100 GeV - 10 TeV scale supersymmetry to achieve
unification of the gauge couplings in view of LEP results.
 (For a thorough review of supersymmetric dark 
matter, see \cite{jkg}.)

When using the minimal supersymmetric standard model 
in calculations
 of relic dark matter density, one should make sure that all
accelerator  constraints on supersymmetric particles and couplings are
imposed. In addition to significant restrictions on parameters given by
LEP \cite{lepbounds}, the measurement of the $b\to s\gamma$ process 
is providing important bounds.

The relic density calculation in the MSSM for a given set of 
parameters is nowadays accurate to 10~\% or so. A recent 
important improvement is the inclusion of 
coannihilations, which can change the relic abundance by a large 
factor in some instances \cite{coann}.

\section{Indirect detection techniques}

In principle, all stable particles produced in annihilation processes in 
the halo can serve 
as signatures of a neutralino dark matter candidate. However, 
electrons and protons are much 
too abundant in the ordinary cosmic rays to be useful. 
Much lower background fluxes, 
and therefore greater potential for detection of an additional 
component, are present in the case of positrons and antiprotons. 
The problem is still that the signal does not have 
a distinct spectral signature (and magnetic fields in the galaxy cause
a diffusion and complete isotropisation of the interstellar flux). 
The featureless energy distribution of positrons and antiprotons 
is due to the fact that the primary 
annihilation processes are into quarks,  heavy leptons, gauge bosons and 
Higgs particles, whereas the positrons 
and antiprotons are secondary or tertiary decay products. 
The reason for the small 
direct coupling to electron-positron pairs is the Majorana nature 
of the neutralino 
coupled with the fact that the annihilation takes place essentially at rest. 
Selection rules 
then force the coupling to fermions to be proportional to the fermion mass. 

There has recently been a new  balloon-borne  detection 
experiment \cite{Barwick}, with increased sensitivity to eventual positrons from neutralino annihilation,
and an excess of positrons over that expected from ordinary sources 
has been found. However, since there are many other possibilities to 
create positrons by astrophysical sources, e.g. in nearby supernova 
remnants, the interpretation is not yet conclusive. However, a new
theoretical calculation has shown that the experimental situation is
compatible with the presence of a neutralino-induced component, if 
the dark matter is clumpy so as to enhance the signal \cite{jetb}.
(Remember that such a non-homogeneous distribution of dark matter is 
in fact what is expected in many cold dark matter formation 
scenarios \cite{moore}.)

 Antiprotons could for some supersymmetric parameters constitute a useful 
signal \cite{antiprotons}, but even with the upcoming space 
experiments \cite{battiston,pamela} it will be quite difficult
to  disentangle a low-energy signal from the smooth 
cosmic-ray induced background.  For kinematical reasons, antiprotons 
created by pair-production in cosmic ray collisions with interstellar 
gas and dust are born with relatively high energy, whereas 
antiprotons from neutralino annihilation populate also the sub-100 
MeV energy band.
A problem that plagues estimates of the signal strength of both positrons and 
antiprotons is, however, the uncertainty of the galactic propagation model 
and solar wind
 modulation. Also, secondary $\bar p$ interactions in the interstellar
medium and $\bar p$ production on helium nuclei also give low-energy
antiprotons which could mask an eventual signal \cite{ourpbar}.

\begin{figure}[!htb]
\centerline{\epsfig{file=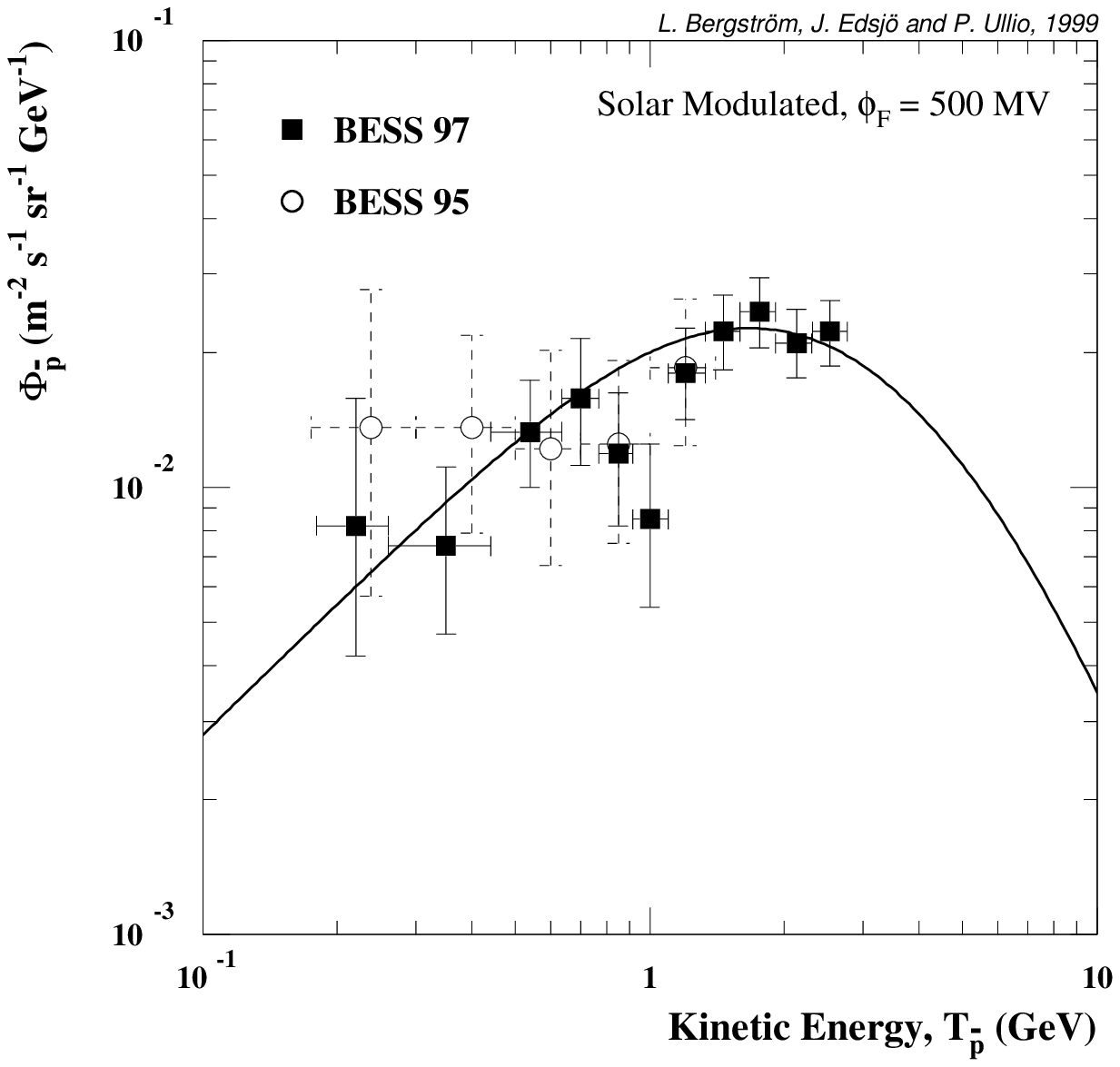,width=0.5\textwidth}
\epsfig{file=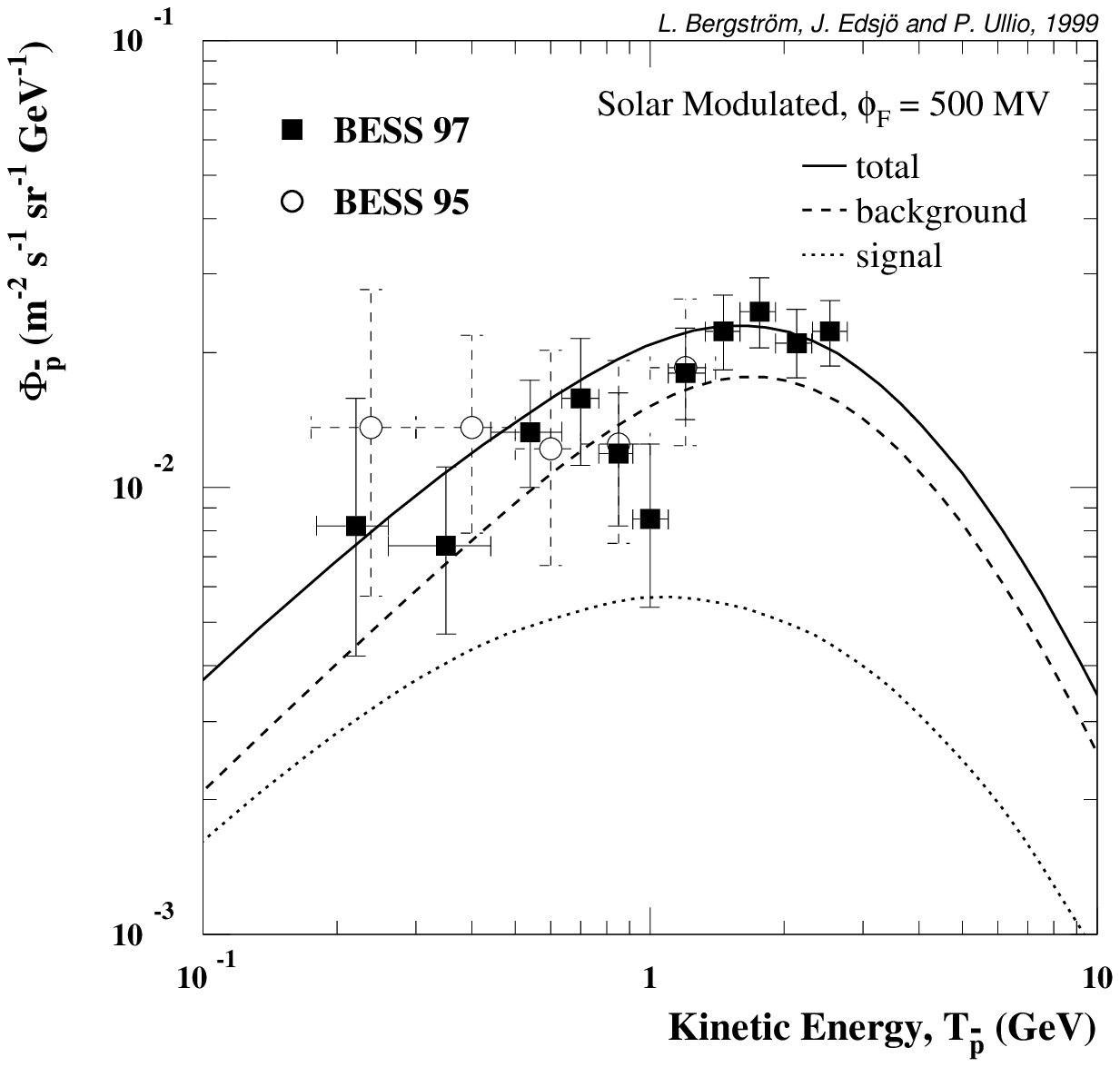,width=0.5\textwidth}
}
\caption{(a) Cosmic-ray antiproton flux as a function of $\bar p$ kinetic
energy. Shown are the BESS 1995 and 1997 data points \protect\cite{bess97}.
The solid line represents the default cosmic-ray induced background
flux according to the analysis in \protect\cite{ourpbar}.
(b) Example of a composite spectrum consisting of the reference
background $\bar p$ flux in (a) reduced by 24 \% with
the addition of the predicted flux from annihilating dark matter neutralinos
of one of the MSSM models of \protect\cite{ourpbar}.}

\label{fig:pbarfig}

\end{figure}

Many supersymmetric models, compatible with all 
 accelerator constraints, give predictions for the 
antiproton flux \cite{bottinonew,clumpylett} 
 which are in agreement with very recent data from the BESS 
 collaboration \cite{bess97}. The question remains, however, whether 
 or not cosmic-ray induced secondary antiprotons can fully explain the
 results. The most complete theoretical calculations \cite{ourpbar,simon} 
 indicate that the band of uncertainties of the conventional flux is 
 large enough to encompass the new data. This is shown in 
 Fig.~\ref{fig:pbarfig} (a), 
where the present BESS data \cite{bess97} are compared to the default
background prediction of \cite{ourpbar}.  In Fig.~\ref{fig:pbarfig} (b)
a 24 \% reduction of the secondary flux (well within the band of 
model uncertainties) is combined with the contribution from one of the 
supersymmetric models
of \cite{ourpbar} to give an equally good fit to the data.

Obviously, it will not be easy to tell these two cases apart,
even with the high-quality data which can soon be expected
\cite{battiston}.

 Even allowing for large  systematic effects, the 
 measured antiproton flux gives, however, rather stringent limits on the 
 lifetime of hypothetical $R$-parity violating decaying 
neutralinos \cite{baltz}.

\subsection{Methods with distinct experimental signature}

With these problems of positrons and antiprotons, one would expect that 
problems
 of gamma rays and neutrinos are similar, if they only arise from 
secondary decays in the 
annihilation process. For instance, the gamma ray spectrum arising from
the fragmentation of fermion and gauge boson final states is quite 
featureless and gives the bulk of the gammas at low energy where the
cosmic gamma ray background is severe. Also, the density of 
neutralinos in the halo is not large enough to give a measurable flux 
of secondary neutrinos, unless the dark matter halo is very clumpy.
 However, neutrinos can escape from the centre  
of the Sun or Earth, where 
neutralinos may have been gravitationally trapped and therefore their density 
enhanced. Also, gamma rays may result from loop-induced 
annihilations \cite{gammaline,zgamma}  
$\chi\chi\to\gamma\gamma$
or $\chi\chi\to Z\gamma$.

The rates of these processes are difficult to estimate because of 
uncertainties in 
the supersymmetric parameters, cross sections and halo density profile. However, 
in contrast to the other proposed detection methods they have 
the virtue of giving  very 
distinct, ``smoking gun'' signals: high-energy neutrinos from the 
centre of the Earth or 
Sun, or monoenergetic photons with $E_\gamma = m_\chi$ or $E_\gamma = m_\chi
(1-m_{Z}^2/4m_{\chi}^2)$ from the halo. In fact, also continuum gammas 
may give a good signature, since the average angular distribution 
should follow the dark matter distribution of the halo, in contrast 
to the background (cosmic-ray induced gamma ray caused mainly by protons 
hitting interstellar hydrogen) which should emanate mainly from the 
disk. This possibility has received increased attention recently due 
to some indications from EGRET data \cite{dixon} that an excess of 
GeV-scale gamma-rays appear to come from the
 halo and could be caused by neutralino 
annihilation \cite{BBU,clumpylett,paololett}. 
However, since there are other 
explanations possible (such as $\pi^0$ production  from cosmic 
rays having escaped the 
disk), it is plausible that other independent evidence (such as 
spectral features - ideally a line) has to be provided before a 
detection of a dark matter-induced signal can be claimed.  

\subsection{Gamma ray lines}

 The detection probability of a gamma line signal depends on
the very poorly known density profile of the dark matter halo.

To illustrate this point, let us consider the characteristic 
angular dependence of 
the gamma-ray intensity from neutralino annihilation in the galactic halo. 
Annihilation of neutralinos in an isothermal halo 
with core radius 
$a$ leads to a gamma-ray flux of
\begin{eqnarray}
     {d{\cal F} \over {d \Omega}}\simeq \,
     (2\times10^{-11} {\rm cm}^{-2} {\rm s}^{-1} {\rm sr}^{-1})\times &
     &\nonumber\\ 
    {(\sigma_{\gamma\gamma} v)_{29}
     (\rho_\chi^{0.3})^2\over (m_\chi/\, 10\,
     {\rm GeV})^2} \,{\left(R\over 8.5\ {\rm kpc}\right)}J(\Psi)&&
\end{eqnarray}
where
 $(\sigma_{\gamma\gamma}
v)_{29}$ is the annihilation rate in units of 
     $10^{-29}\, {\rm cm}^3\,{\rm s}^{-1}$,
$\rho_\chi^{0.3}$ is the
local neutralino halo density in units of 0.3 GeV cm$^{-3}$  
and $R$ is the distance 
to the galactic center.
The integral $J(\Psi)$ is given by
\beq
J(\Psi)={1\over R\rho_{0}^2}\int_{\rm 
line-of-sight}\rho^2(\ell)d\ell(\Psi),
\eeq
and is evidently very  sensitive to local density variations along the 
line-of-sight path of integration.

 We remind of the 
fact that since the 
neutralino velocities in the halo are of the order of 10$^{-3}$ of the 
velocity of light, the 
annihilation can be considered to be at rest. The resulting gamma ray 
spectrum is a line 
at $E_\gamma=m_\chi$ of relative linewidth 10$^{-3}$ which in 
favourable cases 
will stand out against background. The process $\chi\chi\to Z\gamma$ 
is treated analogously and has a  similar rate \cite{zgamma}.

To compute $J(\Psi)$, a model of the dark matter halo has to be chosen. 
Recently, N-body simulations have given a clue to the final halo 
profile obtained by hierarchical clustering in a CDM scenario \cite{NFW}.
It turns out that the universal halo profile found in 
these simulations has a rather significant enhancement $\propto 1/r$ 
near the halo centre. If applicable to the Milky Way, this
 would lead to a much enhanced annihilation 
rate towards the galactic centre, and also to a very characteristic 
angular dependence of the line signal. This would be very beneficial 
when discriminating against the galactic and extragalactic $\gamma$ 
ray background, and Air Cherenkov Telescopes (ACTs) would be eminently 
suited to look for these signals, if the energy resolution is at the 
$10-20$ \% level.

\begin{figure}[!htb]
\centerline{\epsfig{file=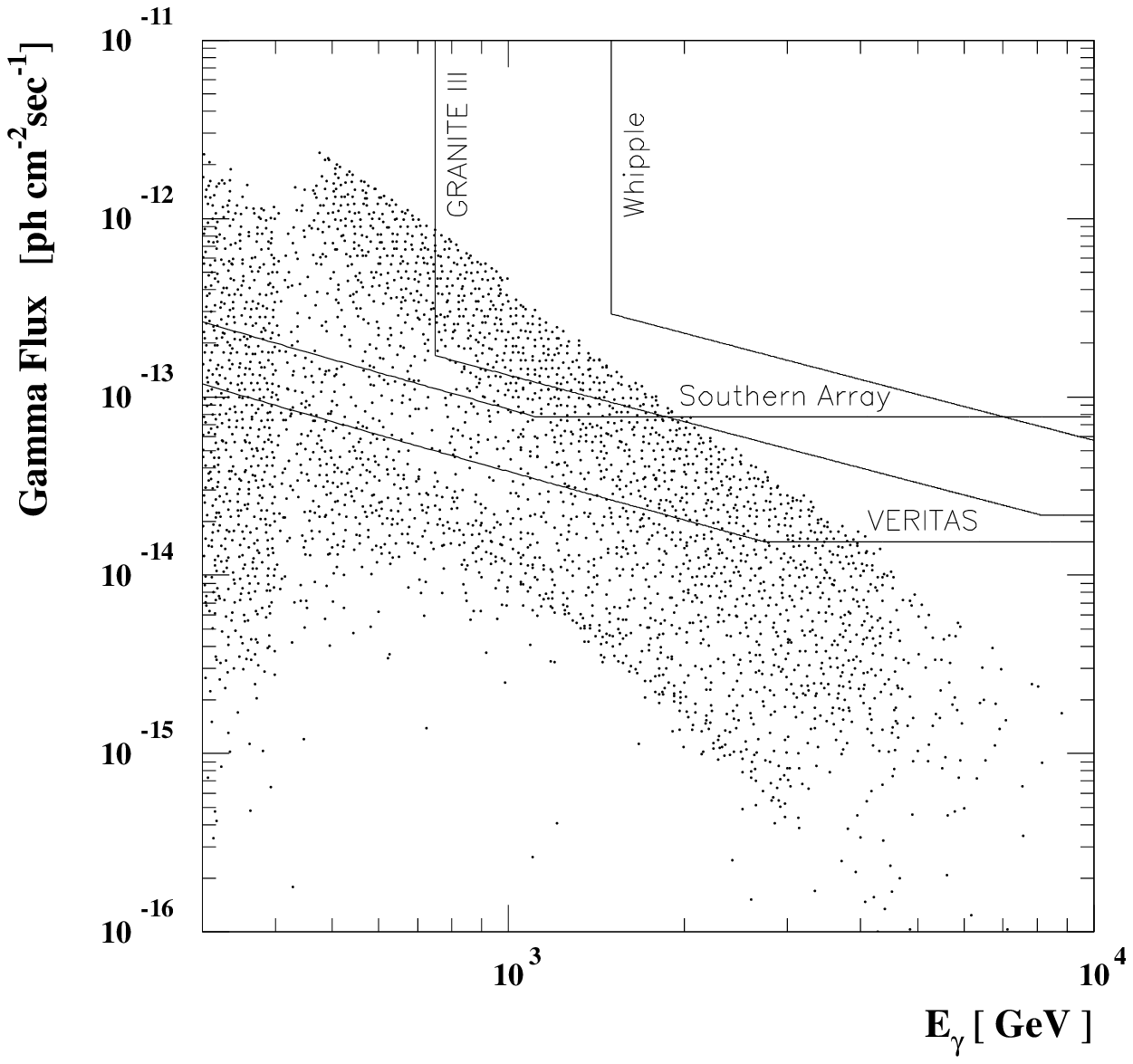,width=0.45\textwidth}\epsfig{file=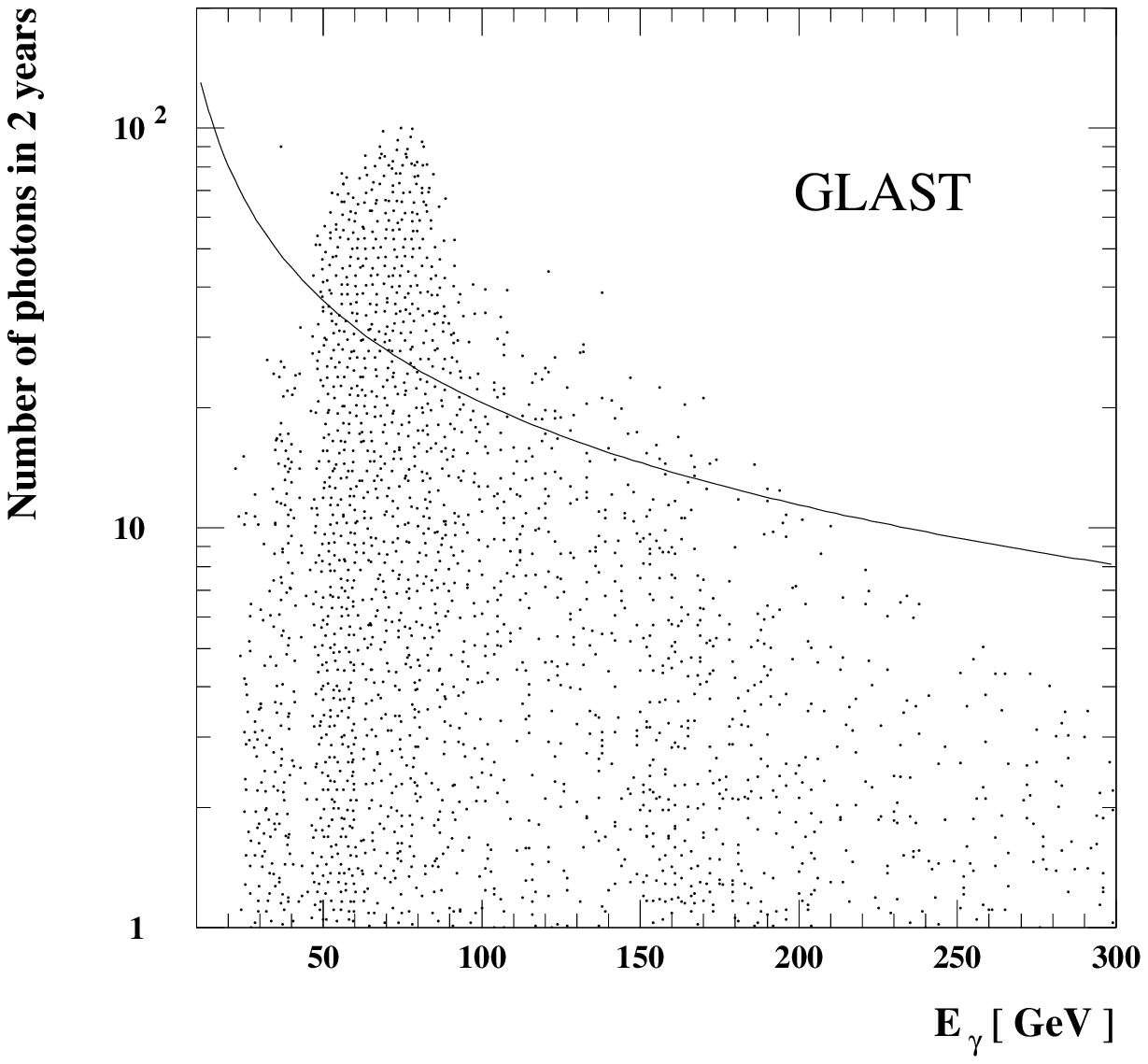,width=0.45\textwidth}}
\caption{Results for the gamma ray line flux in an extensive 
scan of supersymmetric 
parameter space in the MSSM \protect\cite{BBU}.
Shown in the figure to the left is the number of events versus photon energy in an Air Cherenkov Telescope of area $5\cdot 10^4$
m$^2$  viewing the galactic centre for one year. The 
halo profile of \protect\cite{NFW} for the dark matter has been assumed.
In the figure to the right the 5-$\sigma$ discovery
limit of the GLAST satellite is shown assuming a 2-year exposure and the
same halo parameters.}
\label{fig:2gamma}
\end{figure}

In Fig.\,~\ref{fig:2gamma}, we show the gamma ray line flux given in a scan
 of supersymmetric
models consistent will all experimental bounds (including $b\to s\gamma$),
assuming an effective value of $10^3$ for the average of $J(\Psi)$ 
over the $10^{-3}$ steradians that typically an Air Cherenkov 
Telescope (ACT) would cover.  (See \cite{BBU} for details.)

It can be seen that the models which give 
the highest rates should be within reach of the new generation of ACTs 
presently being constructed. These will have an effective area of almost $10^5$
m$^2$, a threshold of some tens of GeV and an energy resolution 
approaching 10 \%. In favourable cases, especially at the low 
$m_{\chi}$ end, also a smaller area detector with  better energy 
resolution and wider angular acceptance such as the proposed GLAST 
satellite \cite{glast} could reach discovery potential, as also shown
in the Figure.

\subsection{Indirect detection through neutrinos}

Another promising indirect detection method is to use neutrinos from 
annihilations of neutralinos accumulated in the centre of the Sun or Earth.
This will be a field of extensive experimental investigations in view
of the new neutrino telescopes ({\sc Amanda}, {\sc Baikal}, {\sc Nestor}, 
{\sc Antares}) 
in operation or under construction.

\begin{figure}[!htb]
\centerline{         \epsfig{file=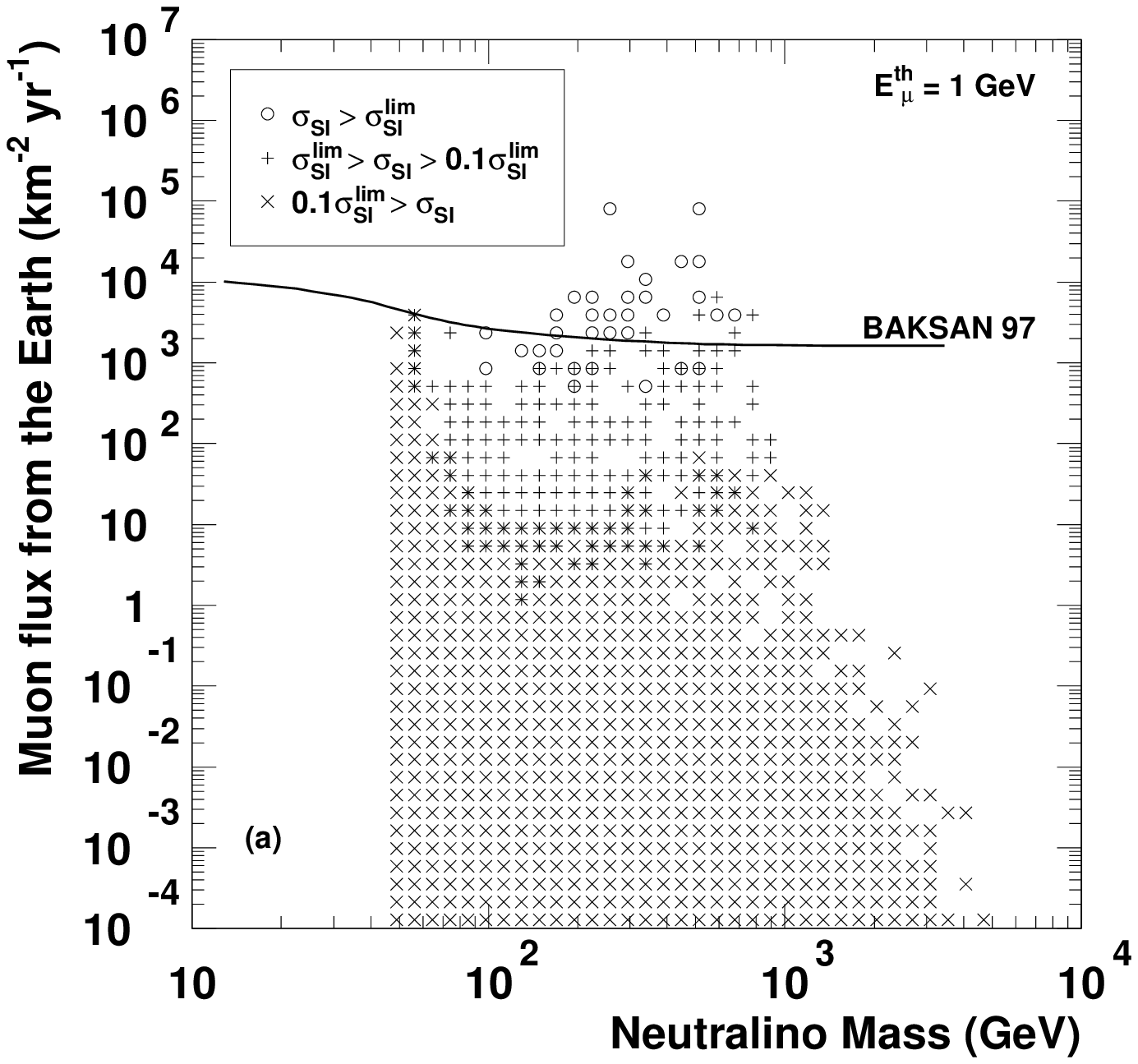,width=0.45\textwidth}         
         \epsfig{file=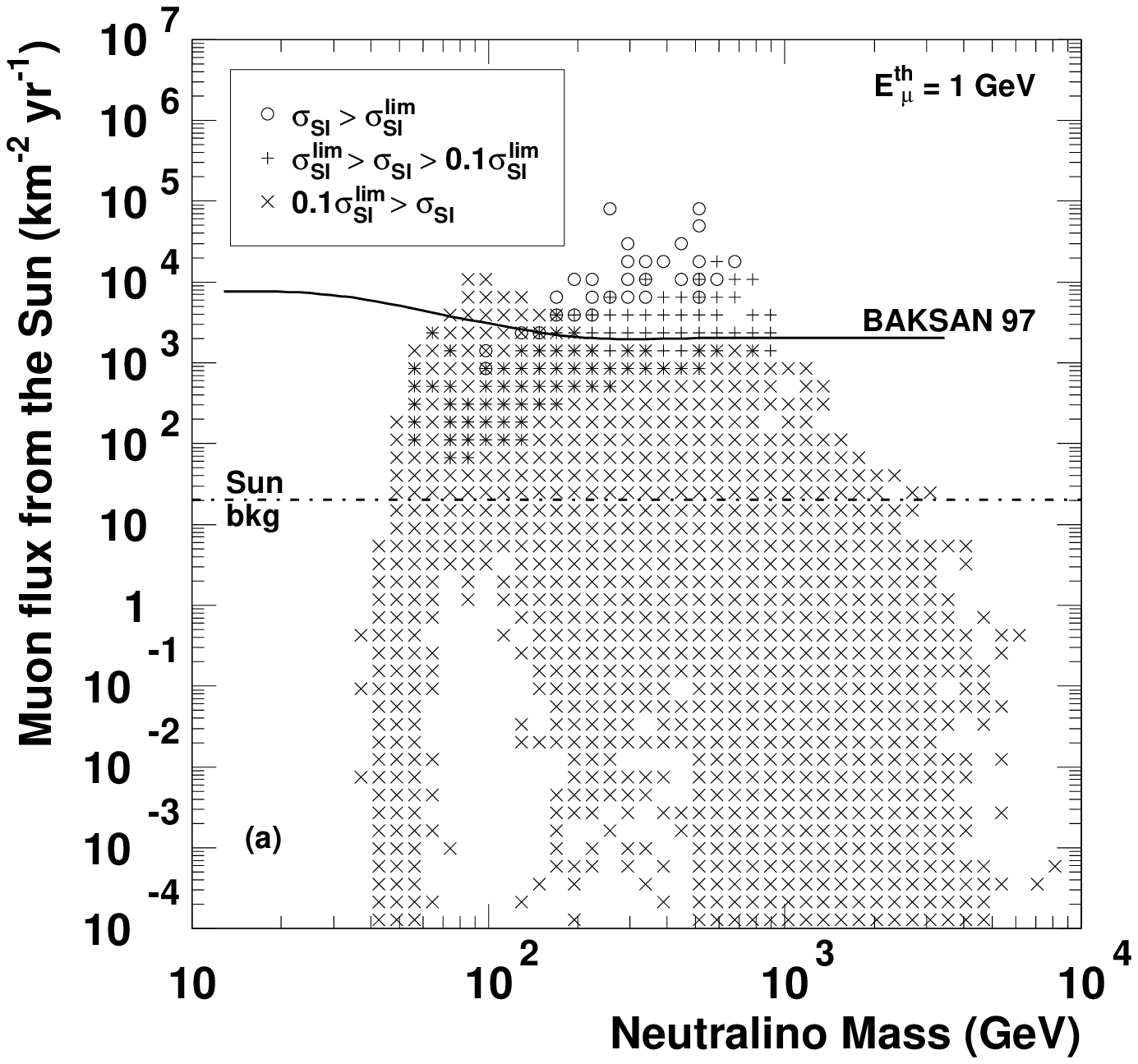,width=0.45\textwidth}}
         \caption{The indirect detection rates from
  neutralino annihilations in the Earth (left) and the Sun (right)
 versus the
  neutralino mass. The
  horizontal line is the Baksan limit \protect\cite{baksan}. For 
  details, see \protect\cite{BEG2}.}
         \label{fig:neutrino}
\end{figure}

To illustrate the potential of neutrino telescopes for discovery of dark matter
through neutrinos from the Earth or Sun, we present the results of a
full calculation \cite{BEG2}. In Fig.\,~\ref{fig:neutrino}
 it can be seen that a neutrino
telescope of area around 1 km$^2$, which is a size currently being discussed,
would have discovery potential for a range of  supersymmetric models.

If a signal were established, one can use the angular spread caused by
the radial distribution of neutralinos (in the Earth) and by the 
energy-dependent mismatch between the direction of the muon and that 
of the neutrino (for both the Sun and the Earth) to get a rather good 
estimate of the neutralino mass \cite{EG}. If muon energy can also be 
measured, one can do even better \cite{BEK}.

It should again be noted  that all indirect detection signals depend 
strongly, and in partly different ways, on the distribution of dark 
matter in the galactic halo. This was investigated in detail in 
\cite{clumpy}, where also a comparison between the various methods 
(including direct detection) was performed. 
\vskip .1cm {\bf Acknowledgements}\vskip .1cm
I wish to thank 
my collaborators, in particular Joakim Edsj\"o, Paolo Gondolo 
and
Piero Ullio, for many
helpful discussions. This work has been supported in part by the
Swedish Natural Science Research Council (NFR).

\end{document}